\documentclass{nature}
\usepackage{graphicx}                           
\usepackage{float}
\usepackage{amsmath}
\usepackage{amsfonts}
\usepackage{amssymb}
\usepackage{comment}
\usepackage{upgreek}
\usepackage[breaklinks,colorlinks,allcolors=blue]{hyperref}
\usepackage{breakurl}
\usepackage{caption}
\newcommand{\co}{$^{12}$CO(2$\rightarrow$1)} 
\newcommand{\hi}{\ifmmode{\rm HI}\else{H\/{\sc i}}\fi} 
\newcommand{\ci}{\ifmmode{\rm CI}\else{C\/{\sc i}}\fi} 
\newcommand{\cii}{\ifmmode{\rm CII}\else{C\/{\sc ii}}\fi} 
\newcommand{\oi}{\ifmmode{\rm OI}\else{O\/{\sc i}}\fi} 
\newcommand{\sone}{MW-C1} 
\newcommand{\stwo}{MW-C2} 
\newcommand{\glon}{\ifmmode{\ell}\else{$\ell$}\fi} 
\newcommand{\glat}{\ifmmode{b}\else{$b$}\fi} 
\newcommand{\vlsr}{\ifmmode{V_\mathrm{LSR}}\else{$V_\mathrm{LSR}$}\fi} 
\newcommand{\vwind}{\ifmmode{V_\mathrm{w}}\else{$V_\mathrm{w}$}\fi} 
\newcommand{\de}{\ifmmode{^\circ}\else{$^\circ$}\fi} 
\newcommand {\kms}{\ifmmode{\rm km \, s^{-1}}\else{$\rm km \, s^{-1}$}\fi} 
\newcommand {\mo}{{\rm M}_\odot}

\newcommand {\moyr}{\,{\rm M_\odot\,\rm yr}^{-1}}

\newcommand{\xco}{X_\mathrm{CO}}

\newcommand{\xcoun}{\mathrm{cm^{-2}} \, (\mathrm{K} \, \kms)^{-1}}

\newcommand{\mhm}{M_\mathrm{mol}}
\newcommand{\mhi}{M_\mathrm{at}}

\title{Cold Gas in the Milky Way's Nuclear Wind}

\author{Enrico M. Di Teodoro$^{1,2,3}$, N.~M. McClure-Griffiths$^2$, Felix J. Lockman$^4$ \& Lucia Armillotta$^{5,2}$}

\begin{document}

\maketitle

\begin{affiliations}
 \item Department of Physics \& Astronomy, Johns Hopkins University, 
Baltimore, MD 21218, USA 
 \item Research School of Astronomy and Astrophysics - The Australian National University, Canberra, ACT, 2611, Australia
 \item Space Telescope Science Institute, Baltimore, MD 21218, USA
 \item Green Bank Observatory, Green Bank, WV 24944, USA
 \item Department of Astrophysical Sciences, Princeton University, Princeton, NJ 08544
\end{affiliations}

\begin{abstract}
The centre of the Milky Way is the site of several high-energy processes that have strongly impacted the inner regions of our Galaxy. 
Activity from the super-massive black hole, Sgr A$^*$, and/or stellar feedback from the inner molecular ring\cite{Molinari+11} expel matter and energy from the disc in the form of a galactic wind\cite{Bland-Hawthorn&Cohen03}. 
Multiphase gas has been observed within this outflow, from hot highly-ionized\cite{Kataoka+13,Ponti+18} ($T\simeq10^6$ K), to warm ionized\cite{Fox+15,Bordoloi+17} ($T\simeq10^{4-5}$ K) and cool atomic gas\cite{McClure-Griffiths+13,DiT+18} ($T\simeq10^{3-4}$ K).
To date, however, there has been no evidence of the cold and dense molecular phase ($T\simeq10-100$ K).
Here we report the first detection of molecular gas outflowing from the centre of our Galaxy. 
This cold material is associated with atomic hydrogen clouds travelling in the nuclear wind\cite{DiT+18}.
The morphology and the kinematics of the molecular gas, resolved on $\sim 1$ pc scale, indicate that these clouds are mixing with the warmer medium and are possibly being disrupted. 
The data also suggest that the mass of molecular gas driven out is not negligible and could impact the rate of star formation in the central regions.
The presence of this cold, dense, high-velocity gas is puzzling, as neither Sgr A$^*$ at its current level of activity, nor star formation in the inner Galaxy seem viable sources for this material. 
%These observations of wind-driven molecular gas clouds open up new possibilities for investigating the detailed mechanisms involved in galactic winds.
\end{abstract}

\begin{figure}
\centering
\includegraphics[width=0.7\textwidth]{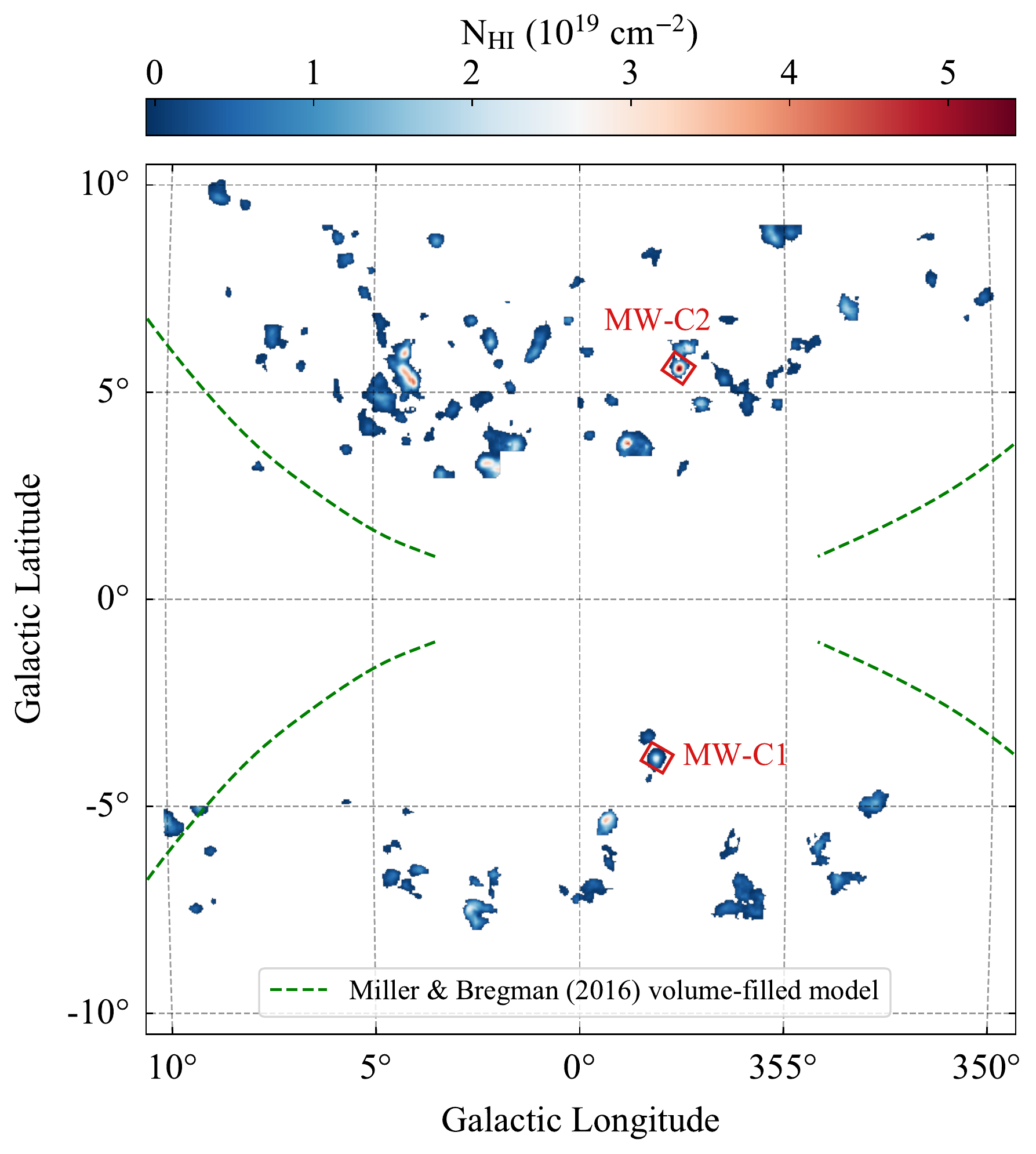}
\caption{Atomic hydrogen gas outflowing from the Galactic Centre. Blue-red colorscale is the colum density of anomalous \hi\ clouds in the MW nuclear wind detected with the GBT\cite{DiT+18}. 
The green-dashed line is the boundary of a volume-filled model for the Fermi Bubbles\cite{Miller&Bregman16}. 
The two \hi\ clouds observed in the \co\ line with APEX are enclosed by red boxes.}
\label{fig:largemap}
\end{figure}

At a distance of only 8.2 kpc\cite{GravityCollaboration+2020}, the Galactic Centre provides a unique laboratory for studying the complex physical processes that occur within a galactic outflow.
The Fermi Bubbles\cite{Bland-Hawthorn&Cohen03,Su+10}, two giant lobes extending up to $\sim 10$ kpc from the Galactic plane, are thought to outline the current boundaries of the Milky Way's nuclear wind.
Several hundred neutral gas clouds have been found recently within this volume through observations of the atomic hydrogen (\hi) line at $\lambda=21$ cm\cite{McClure-Griffiths+13,DiT+18}. 
Figure~\ref{fig:largemap} shows a column density map of \hi\ clouds in the nuclear wind\cite{DiT+18} detected with the Green Bank Telescope (GBT). 
Although the bulk of the cloud population lies within the boundaries of the Fermi Bubbles (green dashed line\cite{Miller&Bregman16}), it has not been established whether this outflowing \hi\ gas arises from the same event that generated the Fermi Bubbles.
These clouds were identified through their anomalous line-of-sight velocities, which are incompatible with Galactic rotation and can instead be described with a bi-conical wind model in which clouds accelerate from the Galactic Centre reaching a maximum velocity of 330 $\kms$ after about 2.5 kpc\cite{DiT+18,Lockman+20}.
To assess whether outflowing \hi\ structures carry molecular gas, we targeted two objects (\sone\ and \stwo, hereinafter), highlighted by red boxes in Figure~\ref{fig:largemap}, in the \co\ emission line at 230.538 GHz with the 12m Atacama Pathfinder Experiment (APEX) telescope. 
These two clouds have relatively high \hi\ column densities ($>10^{19}$ cm$^{-2}$) and show an elongated head-to-tail morphology along the direction pointing away from the Galactic Centre.
We mapped both clouds in \co\ emission over a $15' \times 15'$ field centred on the peak of the \hi\ emission, at a spatial resolution of $28''$ (Full Width at Half Maximum, FWHM), corresponding to $\sim1$ pc at the distance of the Galactic Centre, and a spectral resolution of 0.25 $\kms$.
These data revealed for the first time molecular gas outflowing from the centre of our Galaxy.

\begin{figure}
\centering
\includegraphics[width=.8\textwidth]{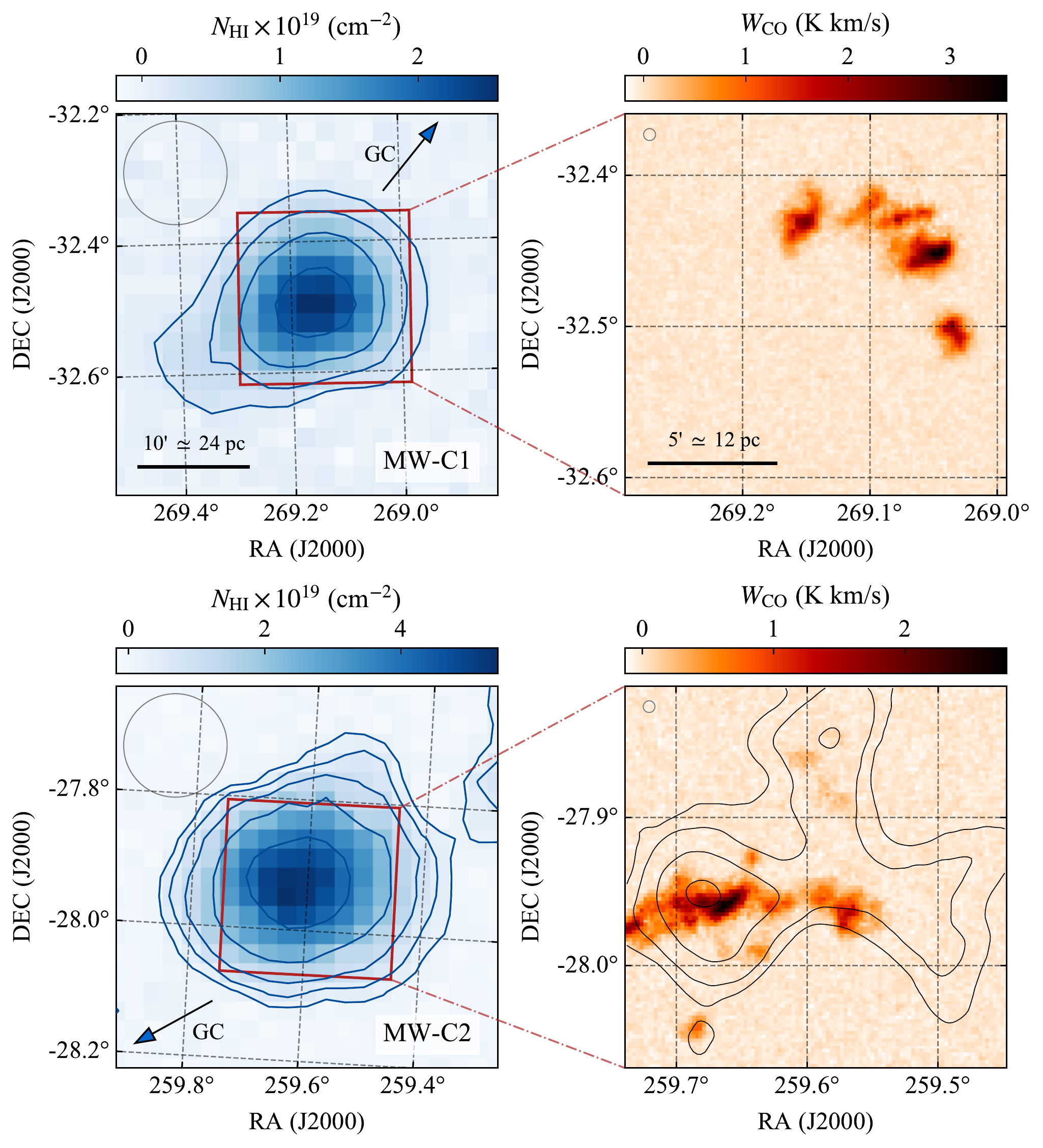} 
\caption{
Atomic hydrogen and molecular gas in two clouds, \sone\ and \stwo\ in the Milky Way nuclear wind. Top panels show MW-C1, bottom panels MW-C2. On the left, \hi\ column density maps from GBT data\cite{DiT+18} at an angular  resolution of $570''$. 
Black arrows point toward the Galactic Centre. 
Red boxes highlight the $15' \times 15'$ fields observed with APEX.
Contour levels are at $(0.2, 0.5, 1, 2, 4) \times 10^{19}$ cm$^{-2}$.
On the right, integrated \co\ brightness temperature maps from APEX data at $28''$ resolution.
\hi\ contours at $(4, 8, 16, 24) \times 10^{20}$ cm$^{-2}$ from ATCA data at $137''$ resolution are overlaid on the \stwo\ map.
Circles at the top-left of each panel show the angular resolution of the telescopes.
}
\label{fig:COdens}
\end{figure}

\begin{figure}
\centering
\includegraphics[width=\textwidth]{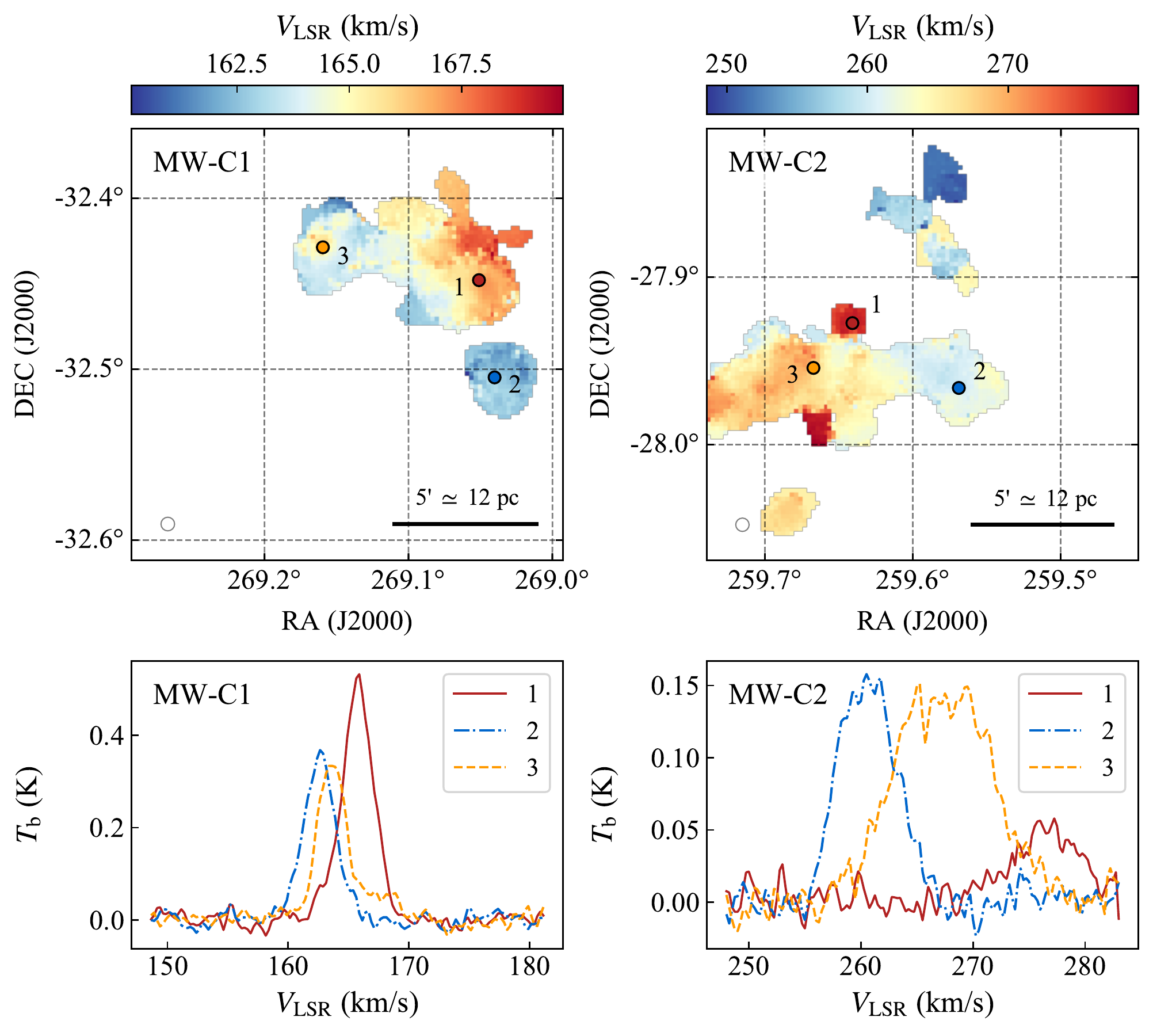}
\caption{Molecular gas kinematics in \sone\ and \stwo\. Left panels show \sone, right panels \stwo. Velocity fields derived through a Gaussian fit to the \co\ data are displayed in the top panels.  Bottom panels show \co\ spectra at the positions labelled on the top panels. Note the difference in velocity spread and line shape between \sone\ and \stwo.}
\label{fig:COvelo}
\end{figure}

Figure~\ref{fig:COdens} shows \hi\ column-density maps (left, blue colorscale) from GBT observations and integrated brightness temperature maps (right, heat colorscale) from the \co\ line with APEX for \sone\ (top) and \stwo\ (bottom).
Higher resolution \hi\ data from the Australia Telescope Compact Array (ATCA) for \stwo\ are overlaid as contours on the CO map.
CO velocity fields and three representative spectra across each field are displayed in Figure~\ref{fig:COvelo}.
CO emission is detected in both \hi\ clouds, with significant morphological and kinematical differences between them. 
\sone\ shows five, distinct, compact clumps of molecular gas concentrated towards the part of the \hi\ cloud that faces the Galactic Centre (arrows in Figure~\ref{fig:COdens}). 
At least three clumps have a velocity gradient along the direction pointing towards the tail of the \hi\ cloud.
All the CO emission in \sone\ lies in the velocity range $\vlsr\simeq160-170 \, \kms$.
Typical FWHM line widths are $\sim2-3 \, \kms$ (see spectra in Figure~\ref{fig:COvelo}). 
In contrast, in \stwo, most of the CO emission is distributed along a filament-like structure, with some fainter and more diffuse clumps in the region away from the Galactic Centre.
CO emission is spread over a larger velocity range than in \sone, spanning 30 $\kms$ over $\vlsr\simeq250-280 \, \kms$, and the velocity field does not show any straightforward ordered motion. 
\co\ line profiles in \stwo\ are much broader than in \sone, with FWHM ranging from $\sim 5-12\, \kms$. 

The observed features indicate that cold gas in \stwo\ is interacting and mixing with the surrounding medium more efficiently than in \sone, resulting in a more turbulent molecular gas.
An interpretation of the differences in the morpho-kinematics of the molecular gas in the two clouds is that we are witnessing two evolutionary stages of a cold cloud being disrupted by the interaction with a hot flow.
Our idealized bi-conical wind model\cite{Lockman+20} with a maximum wind velocity of $330\,\kms$ places \sone\ at a distance of 0.8 kpc and \stwo\ at a distance of 1.8 kpc from the Galactic Centre, implying that \stwo\ may have been within the nuclear outflow twice as long (7 Myr vs 3 Myr). 
Our model also predicts that \stwo\ is moving faster than \sone\ ($\sim300 \, \kms$ vs $\sim240 \,\kms$).
In the classical picture where cold gas is entrained in the hot wind, \sone\ may therefore represent an early stage of the interaction with the surrounding medium, where molecular gas is still relatively intact and undisturbed near the initial dense core, while molecular gas in \stwo\ could have been stripped off from its core, resulting in a disordered morphology/velocity field and broader linewidths.
However, the observed characteristics of the two clouds may also be explained in terms of different local conditions of the hot outflow. 
A larger and more complete sample of molecular gas detections in outflowing clouds is needed to have a more robust picture.

The two clouds analysed in this work have atomic gas masses of $\mhi\simeq220\,\mo$ (\sone) and $\mhi\simeq800\,\mo$ (\stwo), as derived from \hi\ data. 
All mass measurements from observations are scaled by a factor 1.36 to account for the presence of Helium.
It is not straightforward to estimate the mass of molecular matter, as the gas may have a significant opacity in the \co\ line, and the appropriate CO-to-H$_2$ conversion factor $\xco$ in  the Milky Way's wind is unknown.
We used the observed CO integrated brightness temperatures, cloud radii and line widths to constrain acceptable $\xco$ values by means of chemical and thermal modelling of a cloud subject to dissociation by photons and cosmic rays.
We found that $\xco$ for the \co\ transition in our clouds lies in the range $\simeq2-40\times10^{20} \, \xcoun$.  
The lowest value, $\xco=2\times10^{20} \, \xcoun$, is consistent with the Galactic conversion factor\cite{Bolatto+13}, and was used to derive lower limits to the molecular gas mass. 
We obtained a $\mhm\gtrsim380\,\mo$ for \sone\ and $\mhm\gtrsim375\,\mo$ for \stwo, implying molecular-to-total gas mass fractions $f_\mathrm{mol} = \mhm/(\mhm+\mhi) \gtrsim0.64$ and $\gtrsim0.32$, respectively.
We emphasise that these values are lower bounds and the molecular gas mass may be higher by a factor of ten.
As a consequence, the total mass of molecular gas in the nuclear wind of the Milky Way is large.
Under the conservative assumption of an average $f_\mathrm{mol} \simeq 0.3-0.5$ for all outflowing \hi\ clouds in the GBT sample, and based on the atomic outflow rate\cite{DiT+18} of $\dot{M}_\mathrm{at}\sim0.1\, \moyr$, we estimated an outflow rate of $\dot{M}_\mathrm{mol} \gtrsim 0.05-0.1\, \moyr$ in molecular gas.
This value is of the same order of magnitude of the star formation rate (SFR) of the Central Molecular Zone\cite{Longmore+13} (CMZ), implying a molecular gas loading factor $\eta = \dot{M}_\mathrm{mol}/\mathrm{SFR}$ at least of the order of unity at a distance $\sim1$ kpc from the Galactic plane, a value similar to that estimated in nearby starburst galaxies\cite{Bolatto+13b}.
This cold outflow affects the gas cycle in the inner Galaxy and may constitute an important mechanism to regulate the star formation activity in the CMZ.

From a theoretical point of view, such a large amount of high-velocity molecular gas is puzzling\cite{Veilleux+20}.
It is believed that cool gas in a disc can be lifted and accelerated by both drag force from a hot outflow\cite{Scannapieco&Bruggen15} and by radiation pressure\cite{Thompson+15}. 
This requires a source of strong thermal feedback and/or radiation feedback.
The Milky Way does not currently have an active galactic nucleus (AGN), nor is the SFR of the inner Galaxy comparable to that of starburst galaxies with known molecular winds (e.g.\ NGC253\cite{Bolatto+13b}).
Current simulations of AGN-driven winds have focussed on very powerful AGN\cite{Mukherjee+16,Richings+18} and there have been no investigations studying whether a relatively small black hole like Sgr A$^*$ could expel large amounts of cold gas, even if it had undergone a period of activity in the recent past.
On the other hand, the current SFR of the CMZ is not large enough to explain the estimated outflow rate of cold gas\cite{Armillotta+19} and no observational evidence to date suggests a significant change in the SFR of the CMZ in the last few Myr\cite{Barnes+17}. 
A scenario where the star formation in the CMZ is episodic on a longer cycle\cite{Krumholz+17,Armillotta+20} ($10-50$ Myr) and is currently near a minimum might help to partly reconcile the observed and predicted cool gas mass loading rates, although our wind model suggests that lifetimes of cold clouds are shorter than 10 Myr.
Cosmic rays are also believed to contribute pressure on cold gas\cite{Girichidis+18}, but their role is only just starting to be understood and needs observational constraints.
Moreover, in either an AGN-driven or a starburst-driven wind, the extent to which cold gas survives under acceleration is a matter of debate\cite{Scannapieco&Bruggen15,Zhang+17}, and several different mechanisms have been investigated to extend the lifetime of cool gas in a hot wind (e.g.,\ magnetic fields\cite{McCourt+15}, thermal conduction\cite{Armillotta+17}).
An alternative scenario has been recently proposed where high-velocity cool neutral gas ($T<10^4$) forms directly within the outflow as a consequence of mixing between slow-moving cool clouds and the fast-moving hot wind\cite{Gronke+18,Schneider+20}.
This mechanism overcomes the problem of accelerating dense material without disrupting it and may explain the high velocities observed in cool outflows. However, current simulations are not able to trace the gas down to the molecular phase.
%However, it is unclear how efficiently molecular hydrogen can be assembled from neutral gas in a hot wind environment of a Milky-Way-like galaxy\cite{Veilleux+20}.

In conclusion, our first detection of ouflowing cold molecular gas in the Milky Way is a challenge for current theories of galactic winds in regular star-forming galaxies, as none of the above processes seems able to easily explain the presence of fast molecular gas in the Milky Way's wind. 
Targeted observations of molecular gas tracers in the Milky Way's nuclear wind promise to contribute significantly to our understanding of these fascinating phenomena.

%%%%%%%%%%%%%%%%%%%%%%%%%%%%%%%%%%%
%%%% METHODS
%%%%%%%%%%%%%%%%%%%%%%%%%%%%%%%%%%%

\newpage
\begin{methods}

\subsection{Observations and data reduction.}
Observations of the \co\ emission line at 230.538 GHz were made with the 12m APEX antenna\cite{Gusten+06} using the PI230 heterodyne receiver (ESO project ID 0104.B-0106A; P.I. Di Teodoro E.~M.). 
The spectrometer\cite{Klein+12} covers a bandwidth of 8 GHz at a spectral resolution of 61 kHz, corresponding to a velocity resolution of about 0.08 $\kms$ at 230 GHz. 
The beam size at this frequency is 27.8'' (FWHM), the main-beam efficiency is 0.72 and the Jy/K conversion factor is $40\pm3$. 
We observed our targets in on-the-fly position-switching mode, integrating for 1 second  every 9$''$.
Both fields were $15' \times 15'$ wide, centred at $(\alpha,\delta)_\mathrm{J2000} = (\mathrm{17h56m34.0s},-32^\circ 29' 14'')$ for \sone\ and at $(\alpha,\delta)_\mathrm{J2000} = (\mathrm{17h18m22.2s},-27^\circ 56' 28'')$ for \stwo.
Observed regions are shown as red boxes in Figure~\ref{fig:largemap} and in Figure~\ref{fig:COdens}.
Total integration time was approximately 25h for each field.
Throughout the observing session (September-November 2019), the precipitable water vapour (PWV) varied between 0.6 and 3 mm.

We reduced the data using the Continuum and Line Analysis Single-dish Software (\textsc{Class}) from the \textsc{Gildas} package\cite{Gildas+13}.
A first-order baseline was subtracted from calibrated spectra by interpolating the channels outside the velocity windows where we expected to see the emission based on the \hi\ observations.
Spectra were then smoothed in velocity and mapped onto a grid with a pixel size of $9''$ and a channel width of 0.25 $\kms$.
The root mean square  noise ($\sigma_\mathrm{rms}$) in the final datacubes is 65 mK and 55 mK in a 0.25 $\kms$ channel for \sone\ and \stwo, respectively.

\subsection{Atomic gas and molecular gas mass.} 
The \hi\ GBT data\cite{DiT+18} and the \co\ APEX data were analyzed to estimate atomic and molecular gas masses, respectively. 
First, the 3D source finder \textsc{Duchamp}\cite{Whiting+12}  
was applied to the data cubes to  identify regions of significant emission. 
During this process, we set a primary threshold to identify emission peaks at $5\times \sigma_\mathrm{rms}$ and  reconstructed sources by adding pixels down to a secondary threshold of $2.5\times \sigma_\mathrm{rms}$, where $\sigma_\mathrm{rms}$ is the rms noise in a given datacube.

The column density at a given position $(x,y)$ on the sky can be written as:

\begin{equation}
N_\mathrm{H}(x,y) =  C \int T_\mathrm{b} (x,y,v) \,dv
\end{equation}

\noindent where the integral considers only pixels in a detection,  $T_\mathrm{b}$ is the line brightness temperature, $dv$ is the channel width (1 $\kms$ for GBT and 0.25 $\kms$ for APEX) and $C$ is a constant.
For the \hi\ line, under the assumption that the gas is optically thin, the constant is\cite{Roberts75} $C= 1.82 \times 10^{18} \, \xcoun$.
For CO lines, this constant is also known as the CO-to-H$_2$ conversion factor $X_\mathrm{CO}$\cite{Bolatto+13}.
Because the conversion factor in the nuclear wind can not be constrained with current data, we used the value estimated in molecular clouds in the MW's disc\cite{Heyer+09}, $X_\mathrm{CO} = 2\times10^{20} \, \xcoun$.
We checked this $\xco$ value against the predictions from radiative transfer models, described in the next Section, and found that the Galactic value is probably a lower limit for clouds in the nuclear wind.
  
The total mass of gas can finally be calculated as:

\begin{equation}
M  = 1.36 \, m \,  D^2 \int N_\mathrm{H}(x,y) \,\,dx\, dy\\
\end{equation}

\noindent where the factor 1.36 takes into account Helium, $D \simeq 8.2$ kpc is the adopted distance to the clouds, $m$ is the mass of atomic/molecular hydrogen for atomic/molecular gas, $dx$ and $dy$ are the pixel sizes in radians ($105''$ for GBT, $9''$ for APEX).
Observed properties and estimated masses are summarized in Extended Data Table~\ref{tab:prop}.

\subsection{Radiative transfer models.} 
We used the chemistry and radiative transfer code \textsc{Despotic}\cite{Krumholz14} to constrain the CO-to-H$_2$ conversion factor of the clouds. \textsc{Despotic} computes the chemical and thermal state of an optically-thick cloud given its
volume density and column density.  The gas turbulent velocity dispersion was assumed to be $1-5 \, \kms$ (see Figure~\ref{fig:COvelo}) in our modelling. The chemical equilibrium calculation uses solar abundances for dust and all elements in the H$-$C$-$O chemical network\cite{Gong+17}, while the thermal equilibrium calculation includes heating by cosmic rays, grain photoelectric effect, cooling by the lines of \hi, \cii, \ci, \oi, and CO, and collisional energy exchange between dust and gas. Level populations were calculated using an escape probability method, with escape probabilities estimated using \textsc{Despotic}'s spherical geometry option.

We investigated different combinations of interstellar radiation field $\chi$ and cosmic ray ionization rate $\zeta$ through a set of \textsc{Despotic} models with $\log (\chi/G_0) = [-1, 0, 1, 2]$, where $G_0$ is the Solar radiation field\cite{Draine78}  and $\log(\zeta/\mathrm{s}^{-1}) = [-16, -15, -14]$. 
The interstellar radiation field  was varied between sub-Solar ($\chi\simeq0.1\, G_0$) and highly super-Solar ($\chi\simeq 100 \,G_0$, a value representative of a highly star-forming environment as the CMZ). 
The cosmic ray ionization rate ranges from the value measured in the Solar neighbourhood\cite{Indriolo+12} ($\zeta\simeq10^{-16} \,  \mathrm{s^{-1}}$) to the estimated upper limit for the CMZ\cite{Oka+19} ($\zeta\simeq10^{-14} \, \mathrm{s^{-1}}$). 
We stress that our CO clouds lie at about 1 kpc from the Galactic Plane and that both the interstellar radiation field and the cosmic ray ionization rate are expected to drop with distance from the disc. Therefore, although the estimated values of $\chi$ and $\zeta$ in the CMZ are orders of magnitude higher than in the Solar neighbourhood, models with intermediate interstellar radiation field and cosmic ray ionization rate should be more representative of conditions high in the Milky Way's wind.

For each model, \textsc{Despotic} returned the \co\ integrated brightness temperatures ($W_\mathrm{CO}$) as a function of number density ($n_\mathrm{H_2}$) and column density ($N_\mathrm{H_2}$) of molecular Hydrogen.
We only considered solutions consistent with the observed integrated brightness temperature ($1-5 \; \mathrm{K} \, \kms$, see Figure~\ref{fig:COdens}) and observed cloud radius $R=0.75n_\mathrm{H_2}/N_\mathrm{H_2}$ ($1-5$ pc) and we calculated the expected CO-to-H$_2$ conversion factor $X_\mathrm{CO}=N_\mathrm{H_2}/W_\mathrm{CO}$ for the \co\ transition.
We found that there are no acceptable solutions for a strong interstellar radiation field ($\log (\chi/G_0) \geq 1$),  indicating that molecular clouds with the observed properties can not exist in the presence of a CMZ-like radiation field. 
Instead, models with Solar and sub-Solar radiation fields returned solutions compatible with the observational constraints for any cosmic ray ionization field. 
An interstellar radiation field weaker than the one produced in the CMZ is therefore more representative of the environment at 1 kpc above the Galactic Centre.
Predicted $X_\mathrm{CO}$ varies by an order of magnitude, ranging between $\simeq 2\times10^{20} \, \xcoun$ and $\simeq 4\times10^{21} \, \xcoun$, depending on the combination of radiation field and cosmic ray ionization rate.
The value of $\xco = 2\times10^{20} \, \xcoun$ commonly assumed in the Milky Way disc\cite{Bolatto+13} and used in this paper is consistent with the smallest values returned by our radiative transfer models, obtained with a weak, sub-Solar, radiation field and a Solar-like cosmic ray ionization rate of $\zeta=10^{-16} \, \mathrm{s^{-1}}$. 
As a consequence, the molecular gas masses calculated in this work likely represent lower limits to the real cold gas mass in our CO clouds.

\subsection{Wind kinematic model.} 
To estimate the position, velocity and lifetime of \sone\ and \stwo, we used a bi-conical wind model\cite{DiT+18,Lockman+20} calibrated on the full population of \hi\ clouds.
This model is based on the assumption that clouds were launched from a small region close to the centre of the Galaxy and are moving with a purely radial velocity $V_\mathrm{w}(r)$, where $r$ is the distance from the Galactic Centre. 
For simplicity, we considered models of the form:

\begin{equation}
    \label{eq:Vw_model}
	V_\mathrm{w} (r) = 
    \begin{cases}
     V_\mathrm{i} + \left( V_\mathrm{max} - V_\mathrm{i} \right)  \frac{r}{r_\mathrm{s}} & {\rm for \,\,\,} r < r_\mathrm{s} \\ 
	 V_\mathrm{max} & {\rm for \,\,\,} r \geq r_\mathrm{s}
    \end{cases}
\end{equation}

\noindent where $V_\mathrm{i}$ is the initial velocity at $r=0$ and $r_\mathrm{s}$ is the scale distance at which the maximum velocity $V_\mathrm{max}$ is reached. 
Eq.~\ref{eq:Vw_model} describes a kinematic model where clouds are subject to a constant acceleration up to $r_\mathrm{s}$ and maintain a constant velocity at distances $r \geq r_\mathrm{s}$. 
Although Eq.~\ref{eq:Vw_model} is purely empirical and chosen to reproduce the \hi\ data, recent hydrodynamical simulations of starburst-driven winds have found qualitatively similar trends for the cool gas velocity with distance\cite{Schneider+20}. 
The local standard of rest (LSR) velocity $V_\mathrm{LSR}$ of a cloud travelling in the wind and seen at Galactic coordinates $(\ell,b)$ can be written as:

\begin{equation}
    V_\mathrm{LSR}(\ell,b,r) =  V_\mathrm{w} (r) (\sin\phi \, \sin b - \cos\phi \, \cos b \, \cos(\ell+\theta)] - V_0 \sin\ell \, \sin b
\end{equation}

\noindent where the polar angle $\phi$ and the azimuthal angle $\theta$ can easily be written as a function of $(\ell,b,r)$\cite{DiT+18} and $V_0=240 \, \kms$ is the rotation velocity of the LSR around the Galactic Centre\cite{Bland-Hawthorn+16}. 
In our model, clouds are restricted inside a bi-cone with half opening angle $\phi_\mathrm{max}$.
We constrained the four free parameters of this model, i.e.\ $V_\mathrm{i}$, $V_\mathrm{max}$, $r_\mathrm{s}$ and $\phi_\mathrm{max}$, by matching the LSR velocity distributions predicted by our model with that observed from the \hi\ cloud population\cite{DiT+18,Lockman+20}. 
Our fiducial model is a bi-conical wind with opening angle $\phi_\mathrm{max}=70^\circ$, where clouds accelerate from an initial velocity $V_\mathrm{i} = 200 \, \kms$ to a maximum velocity $V_\mathrm{max} = 330 \, \kms$ at $r_\mathrm{s} = 2.5$ kpc\cite{Lockman+20}.
According to this wind model, \sone\ and \stwo\ have travelled a distance of 0.8 kpc and 1.8 kpc from the Galactic Centre in about 3 Myr and 7 Myr, and their current outflow velocity is about 240 $\kms$ and 300 $\kms$, respectively.

\end{methods}

%%
%% TABLES
%%
%% If there are any tables, put them here.
%%

\begin{table*}[t]
\centering
\caption{\small Measured and derived properties of molecular gas clouds outflowing from the Milky Way.
We list: Galactic coordinates ($\ell$, $b$); height from the Galactic plane ($z$); distance from the Galactic Centre ($r$) from our bi-conical outflow model\cite{Lockman+20}; peak \co\ brightness temperature ($T_\mathrm{b,peak}$); typical CO line widths (FWHM); velocity range of CO line in local standard of rest (LSR); lower limits to molecular masses ($\mhm$), derived from \co\ data; atomic gas masses ($\mhi$), derived from \hi\ data. Masses include Helium.}
\label{tab:prop}
\begin{tabular}{lccccccccc}
\noalign{\vspace{5pt}}\hline\hline\noalign{\vspace{5pt}}
 & $\ell$ & $b$ & $z$ & $r$ & $T_\mathrm{b,peak}$ & FWHM & Vel. range & $\mhm$ & $\mhi$ \\
 & ($^\circ$) & ($^\circ$) & (kpc) & (kpc) & (K) & ($\kms$) &  ($\kms$) & ($\mo$) & ($\mo$) \\
\noalign{\vspace{5pt}}\hline\hline\noalign{\vspace{5pt}}
MW-C1 & 358.14 & -3.84 & 0.6 & 0.8 & 1.5 & $2-3$  & $160-170$ & 380 & 220 \\
MW-C2 & 357.58 & 5.56 & 0.9 & 1.8  & 0.5 & $5-12$ & $250-280$ & 375 & 800 \\
\noalign{\vspace{5pt}}\hline\noalign{\vspace{5pt}}
\end{tabular}
\end{table*}

%% Here is the endmatter stuff: Supplementary Info, etc.
%% Use \item's to separate, default label is "Acknowledgements"

\begin{addendum}
 \item[Acknowledgements] 

E.D.T. and L.A. thank E. Ostriker, C.-G. Kim and J.-G. Kim for useful discussions, and M. Krumholz for support with the \textsc{Despotic} code.
E.D.T. was supported by the National Science Foundation under grant 1616177.
E.D.T. and N.M.-G. acknowledge the support of the Australian Research Council (ARC) through grant DP160100723. N.M.-G. acknowledges funding from the ARC via Future Fellowship FT150100024. 
CO observations were made with APEX under ESO proposal 0104.B-0106A.
APEX is a collaboration between the Max-Planck-Institut fur Radioastronomie, the European Southern Observatory, and the Onsala Space Observatory. 
The Green Bank Observatory is a facility of the National Science Foundation operated under a cooperative agreement by Associated Universities, Inc.
The Australia Telescope Compact Array is part of the Australia Telescope National Facility which is funded by the Australian Government for operation as a National Facility managed by CSIRO.

\item[Author contribution] 
E.D.T., N.Mc.-G.\ and F.J.L.\ developed the idea for the project. 
E.D.T.\ reduced and analysed the APEX data, L.A.\ ran the radiative transfer models.  E.D.T.\ wrote the paper with direct contributions from N.Mc.-G., F.J.L.\ and L.A.  All authors reviewed the manuscript.

 \item[Data availability] 
 The APEX raw datasets analysed during the current study will be available at the end of the proprietary period (September 2020) on the ESO archive: \href{http://archive.eso.org/eso/eso_archive_main.html}{http://archive.eso.org/eso/eso\_archive\_main.html}. 
The GBT raw datasets are publicly available at NRAO archive: \href{https://science.nrao.edu/facilities/gbt/software-and-tools}{https://science.nrao.edu/facilities/gbt/software-and-tools}.
The corresponding author will also provide fully-reduced data on reasonable request. 

 \item[Code availability] Software used in this work is publicly available. The \textsc{Gildas/Class} packages for sub-mm data reduction can be found at \href{https://www.iram.fr/IRAMFR/GILDAS/}{https://www.iram.fr/IRAMFR/GILDAS}. \textsc{Duchamp} source finder can be downloaded from \href{https://www.atnf.csiro.au/people/Matthew.Whiting/Duchamp/}{https://www.atnf.csiro.au/people/Matthew.Whiting/Duchamp}.
\textsc{Despotic} radiative transfer code is available at \href{https://bitbucket.org/krumholz/despotic/src/master/}{https://bitbucket.org/krumholz/despotic}.

 \item[Competing Interests] The authors declare that they have no
competing financial interests.
 
 \item[Correspondence] Correspondence and requests for materials
should be addressed to Enrico Di Teodoro 
%(email: \href{mailto:editeodoro@jhu.edu}{editeodoro@jhu.edu}).

\end{addendum}

%% Put the bibliography here, most people will use BiBTeX in
%% which case the environment below should be replaced with
%% the \bibliography{} command.

\section*{References} 
\vspace*{1cm}
\bibliographystyle{naturemag}
\bibliography{MWCO_biblio.bib}

\end{document}